\documentclass[prl,twocolumn,amsmath,amssymb,amsfonts]{revtex4}%
\usepackage{graphicx}
\usepackage{dcolumn}
\usepackage{bm}
\usepackage{amsmath}
\usepackage{amsfonts}
\usepackage{amssymb}%
\begin{document}
\title{Point-contact tunneling involving low-dimensional spin-triplet superconductors}
\author{C.~J.~Bolech}
\author{T.~Giamarchi}
\affiliation{Universit\'{e} de Gen\`{e}ve, DPMC, 24 Quai Ernest Ansermet, CH-1211
Gen\`{e}ve 4, Switzerland}
\date{July 4$^{\text{th}}$, 2003}

\begin{abstract}
We modify and extend previous microscopic calculations of tunneling in
superconducting junctions based on a non-equilibrium Green function formalism
to include the case of spin-triplet pairing. We show that distinctive features
are present in the I-V characteristics of different kinds of junctions, in
particular when the effects of magnetic fields are taken into account, that
permit to identify the type of pairing. We discuss the relevance of these
results in the context of quasi one-dimensional organic superconductors like
\textrm{(TMTSF)}$_{2}$\textrm{PF}$_{6}$ and layered compounds like
\textrm{Sr}$_{2}$\textrm{RuO}$_{4}$.

\end{abstract}
\maketitle

Unconventional superconductivity is a generic classification for all those
superconducting materials that do not fit into the \textit{s}-wave
spin-singlet pairing picture of standard BCS theory\cite{sigrist1991}. One of
the possible extensions of that theory considers anisotropic pairing
mechanisms accompanied by spin-triplet states. One such unconventional
scenario is realized in the \textit{p}-wave spin-triplet superfluid state of
$^{3}$\textrm{He}. Among the superconductors the appearance of unconventional
superconductivity, always accompanied by strong correlations, results in most
cases in spin-singlet pairing. The search for triplet pairing
superconductivity in strongly correlated electron systems becomes thus
natural. It was proposed\cite{rice1995}, for instance, in the case of
\textrm{Sr}$_{2}$\textrm{RuO}$_{4}$ and has to date a considerable amount of
experimental support\cite{mackenzie2003}. This compound shows, in common with
the high-\textit{T}$_{\mathrm{c}}$ cuprates, a layered structure. Other
candidates for triplet pairing, though the evidence is less conclusive, are in
the class of so called organic superconductors\cite{ishiguro2002plus} like
f.i.~\textrm{(TMTSF)}$_{2}$\textrm{X}, with \textrm{X} a counter anion as
\textrm{PF}$_{6}$, \textrm{ClO}$_{4}$, etc. Materials of this class are
low-dimensional conductors in their normal state.

A distinctive feature of triplet superconductors is the possibility of large
upper critical fields; this is contrary to BCS superconductors for which the
paramagnetic response of the normal state puts a limit to the field up to
which the superconducting phase is energetically favorable even in the absence
of orbital effects (the Clogston or BCS Pauli paramagnetic limit,
$H_{\mathrm{p}}$). It was indeed observed during the last few years that at
low temperatures the above mentioned organic compounds show upper critical
fields that exceed $H_{\mathrm{p}}$, usually depending on the angular
orientation of the field with respect to the crystalline
axes\cite{ishiguro2002plus}. In general, for layered and chained compounds,
superconductivity at high fields is attained when the field is oriented so
that orbital effects are suppressed. Although high critical fields constitute
an indication of triplet-pairing, unambiguous interpretation of this effect is
difficult because one needs to rule out other possible phases able to
accommodate large magnetic fields (\textit{e.g.} the
Larkin-Ovchinikov--Fulde-Ferrel phase)\cite{lebed1999}.

It would be highly desirable to have a direct indication of the nature of the
pairing, as in the case of cuprates. A possible experimental determination of
the symmetry of the pairing could come from tunneling experiments. In the past
some of the most crucial experimental verifications of BCS theory came from
tunneling experiments; and, after the discovery of the Josephson effect, some
of the most important practical applications of superconductivity involve
tunneling junctions. In addition, tunneling \textit{per se} acquired renewed
relevance with the development of scanning tunneling microscopy (STM),
actively used nowadays in the study of superconductivity. The simplest
theoretical models used to interpret tunneling experiments in superconductors
are those that go under the name of \textit{semiconducting band models}%
\cite{blonder1982,octavio1983}. A more systematic approach is that based on
tunneling Hamiltonians\cite{cohen1962,wilkins1969,cuevas1996}. Two limiting
cases are usually considered in the calculations: planar interfaces and point
contacts (the latter usually realized in experiments using break junctions,
pressed crossed wires or close STM contacts).

In this letter we focus our attention on the point-contact case. We extend
previous tunneling Hamiltonian calculations to include the case of
triplet-pairing superconductors. As compared to previous calculations, we
simplify considerably the formalism making it more versatile and easy to
implement. Our aim is to explore the possible use of tunneling experiments as
a way of characterizing the type of pairing in unconventional superconductors.
Using non-equilibrium Keldysh Green functions we calculate the full
current-voltage characteristics of different types of tunnel junctions mixing
normal metals, singlet and triplet superconductors. We also study the effects
of applied external magnetic fields on the transport properties of the
different junctions.

Our starting point is the tunneling Hamiltonian,
\begin{equation}
\left.
\begin{array}
[c]{l}%
H_{\mathrm{jun}}=H_{1}+H_{2}+H_{\mathrm{tun}}\\
H_{\mathrm{tun}}=-t~\sum_{\sigma}\left[  \psi_{2,\sigma}^{\dagger}\left(
x=0\right)  \psi_{1,\sigma}^{%
\phantom{\dagger}%
}\left(  x=0\right)  +\mathrm{h.c.}\right]
\end{array}
\right.  \label{eq:hamjun}%
\end{equation}
The first two terms describe the two leads (superconducting or otherwise) and
the third one models the tunneling processes in which an electron with spin
$\sigma$ hops from one lead into the other. The current is proportional to the
rate of change in the relative particle number, $\mathrm{I}=\frac{e}%
{2}\left\langle \partial_{t}\left(  N_{2}-N_{1}\right)  \right\rangle
=\frac{e}{2i}\left\langle \left[  H_{\mathrm{tun}},N_{1}-N_{2}\right]
\right\rangle $.

To model the leads in calculations of point-contact transport on conventional
superconductors, simple models suffice to achieve quantitative agreement with
the experiment. Since dimensionality plays no role in the tunneling, contrary
to the case in some planar junction experiments, all the standard calculations
can be formulated using one-dimensional leads. The situation is more complex
for unconventional superconductors, for which, the anisotropic nature of the
pairing has to be taken into account. However, the ruthenates and organic
superconductors are deemed to have \textit{p}-wave symmetry, and since both
\textit{s}-wave and \textit{p}-wave symmetries can be modeled in a one
dimensional chain, we can set up a formalism that comprises both cases. We
consider a one dimensional band with two Fermi points and expand the fermion
fields around them defining left ($L$) and right ($R$) moving fields. Using
these fields, and in the spirit of BCS theory, we introduce the four gap
functions:
\begin{equation}
\Delta_{a}\left(  x\right)  \propto\left\langle \alpha~\psi_{L\bar{\alpha}}^{%
\phantom{\dagger}%
}\left(  x\right)  ~\sigma_{\alpha\beta}^{a}~\psi_{R\beta}^{%
\phantom{\dagger}%
}\left(  x\right)  \right\rangle \label{eq:delta}%
\end{equation}
where $\alpha,\beta\in\left(  \downarrow,\uparrow\right)  \equiv\left(
-1,+1\right)  $ are summed over ($\bar{\alpha}\equiv-\alpha$); $a=0,\ldots,3$;
$\sigma_{\alpha\beta}^{0}$ is the identity matrix while the other three are
the usual Pauli matrices. With this definition $\Delta_{0}\left(  x\right)  $
is the conventional spin-singlet order parameter and the other three gap
functions form a vector of spin-triplet order parameters\cite{mackenzie2003},
$\vec{\Delta}\left(  x\right)  =\Delta\left(  x\right)  \mathbf{\hat{d}%
}\left(  x\right)  $. Approximating the order parameter to have no spatial
dependence, we write in Fourier space
\[
K_{n}=\xi_{ck\alpha}^{n}\psi_{ck\alpha}^{\dagger}\psi_{ck\alpha}^{%
\phantom{\dagger}%
}-\left[  \Delta_{a}\left(  \psi_{Rk\beta}^{\dagger}~\sigma_{\beta\alpha}%
^{a}~\alpha~\psi_{L\bar{k}\bar{\alpha}}^{\dagger}\right)  +\mathrm{h.c.}%
\right]
\]
where $K_{n}=H_{n}-\mu_{n}N_{n}$ with $\mu_{n}$ the chemical potential of lead
$n$. All the indexes are summed over, in particular $c\in\left(  L,R\right)
\equiv\left(  -1,+1\right)  $ sums over the two possible chiralities and
$\xi_{ck\alpha}^{n}=ck-\mu_{n}-\alpha h$ are the corresponding linear
dispersions, shifted by the inclusion of chemical potential and magnetic field
along the $\hat{z}$-axis. We take the quantization axis (\thinspace$\hat{z}%
$\thinspace) along the field direction and consider the cases of triplet order
parameters parallel or perpendicular to it.

Since we are dealing with an out of equilibrium phenomenon, we use Keldysh
formalism\cite{keldysh1965} to treat the tunneling term to all orders,
calculating the full I-V line and giving a quantitative account of its subgap
structure. One past implementation of such an approach, for the \textit{s}%
-wave case, reduced the problem to the solution of a set of linear recursion
relations\cite{cuevas1996}. Here, instead, we notice that in this formalism
the current becomes%
\begin{equation}
I=\frac{et}{2i}~\sum_{\sigma}\int\frac{d\omega}{2\pi}\left\langle
\psi_{2,\sigma}^{\dagger}\psi_{1,\sigma}^{%
\phantom{\dagger}%
}-\psi_{1,\sigma}^{\dagger}\psi_{2,\sigma}^{%
\phantom{\dagger}%
}\right\rangle _{\mathrm{kel}} \label{eq:curkel}%
\end{equation}
where `kel' denotes the Keldysh component of the correlation function and the
$\psi$ stand for $\psi(\omega,x=0)$. Since the current depends only on the
fields at $x=0$ one can analytically integrate the $x$ dependence in the leads
to obtain from Eq.~(\ref{eq:hamjun}) a \textit{local}\emph{ }and\emph{
}\textit{quadratic} Keldysh action for the fields at $x=0$: $S_{\mathrm{jun}%
}=S_{1}+S_{2}+S_{\mathrm{tun}}$. Here $S_{\mathrm{tun}}$ is obtained from
$H_{\mathrm{tun}}$ and $S_{n}$ is the local action of lead $n$, of the form
$S_{n}=\int\frac{d\omega}{2\pi}\mathbf{\Psi}_{n}^{\dagger}(\omega)\hat{g}%
^{-1}\mathbf{\Psi}_{n}^{%
\phantom{\dagger}%
}(\omega)$. Where $\mathbf{\Psi}_{n}$ is an $8$ component spinor (a Keldysh
extended Nambu-Eliashberg spinor), and $\hat{g}^{-1}$ is a matrix whose
components can be computed from $K_{n}$. Its inverse ($\hat{g}$) is given by
the standard advanced, retarded and Keldysh components of the local Green
functions of the lead. As an example we give the expressions for the case when
$\Delta_{1}=\Delta_{2}=0$,%
\begin{align*}
g_{c\sigma,c\sigma}^{\left[  \mathrm{ret,adv}\right]  }  &  =\frac{-\left(
\omega-\mu_{n}+c\sigma h\pm i0^{+}\right)  }{2\sqrt{\left\vert \Delta
_{0}+c\sigma\Delta_{3}\right\vert ^{2}-\left(  \omega-\mu_{n}+c\sigma h\pm
i0^{+}\right)  ^{2}}}\\
g_{c\sigma,\bar{c}\bar{\sigma}}^{\left[  \mathrm{ret,adv}\right]  }  &
=\frac{\left(  \Delta_{0}+c\sigma\Delta_{3}\right)  ^{\left[  \ast\right]
_{c=L}}}{2\sqrt{\left\vert \Delta_{0}+c\sigma\Delta_{3}\right\vert
^{2}-\left(  \omega-\mu_{n}+c\sigma h\pm i0^{+}\right)  ^{2}}}%
\end{align*}
The notation in the numerator of the anomalous functions means that complex
conjugation is in order when $c=L$. The Keldysh component is obtained
immediately as $g^{\mathrm{kel}}=\left(  g^{\mathrm{ret}}-g^{\mathrm{adv}%
}\right)  \tanh\left(  \left(  \omega-\mu_{n}\right)  /2T\right)  $.

Since the action is quadratic, it can be diagonalized which allows to compute
the current from (\ref{eq:curkel}). For normal leads this can be done
analytically. However, for superconducting leads, special attention must be
paid to the fact that frequencies have different reference Fermi levels in
each lead when there is a bias applied. Within each lead, frequencies with
equal positive and negative shifts from the Fermi level form part of the same
spinor element and the corresponding states are related by the coherent
pairing processes in the superconductors; across leads, same frequency states
are related by the tunneling matrix elements of the action. As a result, the
full action for the junction is no longer frequency diagonal. However, since
it is quadratic it can still be written explicitly as a matrix. To each value
($\omega_{0}$) in the frequency window (of size $eV=\mu_{1}-\mu_{2}$) defined
by the chemical potentials in the two leads, one infinite set of related
frequencies can thus be assigned ($p>0$):
\begin{equation}
\left\{
\begin{array}
[c]{r}%
\omega_{p}=2\mu_{2-p\operatorname{mod}2}-\omega_{p-1}\\
\omega_{-p}=2\mu_{1+p\operatorname{mod}2}-\omega_{1-p}%
\end{array}
\right.  \label{q:freq}%
\end{equation}
These sets are all independent and the action is block diagonal between
different ones. Discretizing the frequencies in this window, automatically
defines a discretization of the whole frequency axis. This allows for a
numerical solution of the problem: we deal with one set of frequencies at a
time, and since the sets are infinite, we truncate their hierarchy at some
distance from the central frequency window. This is equivalent to introducing
a \textit{soft} limit in the number of allowed Andreev reflections
($N_{\mathrm{A}}$). We can then numerically invert the corresponding
block-diagonal action, written as a matrix in frequency space. The
off-diagonal Green function matrix elements thus obtained allow to compute the
current using Eq.~(\ref{eq:curkel}).

The practical implementation of this approach is quite simple and allows to
consider the (combined) effects of finite temperature, applied magnetic
fields, contact potentials in the junction, spin-flip tunneling or spin-flip
scattering processes in the leads. It is also possible to compute the
a.c.~response. Our interest is in comparing singlet and triplet superconductor
junctions and how they respond differently in the presence of an external
field; we will ignore other additional complications.

\begin{figure}
[t]
\begin{center}
\includegraphics[
height=3.5215in,
width=5.028in
]%
{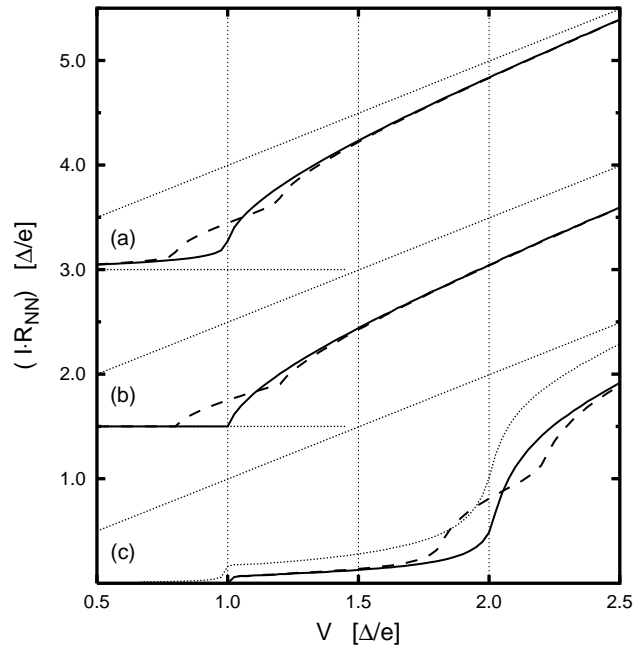}%
\caption{Three different sets of I-V characteristics vertically displaced for
clarity. From top to bottom: (a) for an N-S junction, (b) N-T junction, (c)
S-T junction. The solid (dashed) lines correspond to the characteristics in
the absence (presence) of an oriented magnetic field (see text for details).
In all three sets is plotted as reference the curve for the N-N case (diagonal
straight dotted lines) and in the third one the curve for an S-S junction is
also given.}%
\label{combinedIV}%
\end{center}
\end{figure}

In order to illustrate the different I-V characteristics for different types
of junctions, we fix a set of parameters. We take $t=0.2$ for the tunneling
overlap integral, and when there is a magnetic field we fix its value to
$h=0.2$ in units of $\Delta$ (the magnitud of the singlet gap, $\Delta_{0}$,
or of the triplet vector order parameter depending on the case). All the
curves we show are for the d.c. response in the limit of vanishing
temperatures. For the truncation procedure we take $N_{\mathrm{A}}=3$ (and
verify that larger values produce identical curves). Let us use the notation,
N: normal-metal, S: singlet-superconductor and T: triplet-superconductor. We
show in Fig.~\ref{combinedIV}\textrm{-(a)} typical curves for an N-S junction.
The effect of the field (for any orientation) is to produce what would be seen
as a Zeeman splitting in the differential conductance peak. Notice the sub-gap
shoulder on the I-V curve when $eV<\Delta$ (for $h=0$); its origin is in the
coherent Andreev processes that take place in the junction contact.%

We show in Fig.~\ref{combinedIV}\textrm{-(b)} typical curves for an N-T
junction. The solid line corresponds to zero field and the dashed line is for
that same junction but in the presence of a field that is aligned with the
vector order parameter. If one considers a field perpendicular to the order
parameter ($\vec{h}\perp\mathbf{\hat{d}}$), the I-V characteristic remains
unaffected, \textit{i.e.} identical to the one for zero field. Notice the
absence of a sub-gap shoulder on the I-V curve, a result of the odd real-space
symmetry of the superconductor probed locally by an ideal point-contact
junction. Given the model we consider, these results would be relevant for
tunneling experiments into a sample edge parallel to the chains, when a
zero-bias peak is not expected \cite{sengupta2001}\footnote{In the presence of
a surface, the $p$-wave order parameter components not perpendicular to it
will not feel its pair breaking effects (remark that in the case of organic
superconductors the order parameter is thought to be strongly anisotropic and
aligned mainly along the $\mathbf{a}$-axis\cite{lebed2000}).}.

Let us now examine the case of S-T junctions, see Fig.~\ref{combinedIV}%
\textrm{-(c)}. One dotted line is the N-N characteristic as in the other
cases; the other is the I-V curve of an S-S junction that shows all the
features well known in the literature\cite{octavio1983,cuevas1996}, in
particular a `sub-gap' shoulder with Andreev steps at $eV=2\Delta/n$ (with
$n=1,2,\ldots$). This curve is, bar orbital effects, not sensitive to applied
fields. The solid line is always the same regardless of the orientation of the
vector order parameter on the triplet-pairing side and the current amplitude
is systematically smaller than in the S-S case. Distinctively, the `sub-gap'
structure shows \textit{only} the first two steps and the current becomes zero
for $eV<\Delta$. Regarding the effect of an applied field, the curve remains
unchanged if the field is parallel to the vector-order-parameter direction,
but shows a Zeeman effect if the field is perpendicular to it (dashed line).
This is to be confronted with the N-T case where the splitting accompanies a
field $\vec{h}\parallel\mathbf{\hat{d}}$.

Let us make contact with the experimental situation on the different compounds
afore mentioned, and dwell first on the particularly interesting case of
\textrm{(TMTSF)}$_{2}$\textrm{PF}$_{6}$. This quasi-one-dimensional organic
compound shows the largest deviations from the $H_{\mathrm{p}}$ limit among
the different members of the Bechgaard salts family, and also as compared with
other organic superconducting salts\cite{ishiguro2002plus}. A recent series of
experiments focused on measuring $H_{\mathrm{c2}}\left(  T\right)  $ for this
material\cite{lee1997,lee2000,lee2002a,lee2002b,oh2004}. The observed large
deviations from the $H_{\mathrm{p}}$ limit as well as the anisotropy and
angular dependence of the upper critical field make the case for an equal-spin
triplet-pairing superconducting phase with an order parameter oriented mainly
along the chains\cite{lebed2000}. Other evidences for spin-triplet pairing in
\textrm{(TMTSF)}$_{2}$\textrm{PF}$_{6}$ also exist, but as with the
$H_{\mathrm{c2}}$ measurements, most of these experiments do not probe the
spin parity directly and their conclusions are difficult to interpret due to
the uncertainties regarding the orbital symmetry of the superconducting phase;
however, recent NMR Knight Shift measurements add strong support to the
spin-triplet scenario\cite{lee2002c}. Here we argue that the magnetic field
response in N-T point-contact tunneling experiments would also constitute a
direct probe of the spin-pairing symmetry (in contrast with tunneling across
planar junctions that is sensitive to the orbital-pairing
symmetry\cite{tanuma2002}). $H_{\mathrm{c2}}$ measurements show that for
magnetic fields along the direction of the conducting chains ($\mathbf{a}%
$-axis) the upper critical field is paramagnetically limited\cite{lebed2000}.
This corresponds, in our convention, to a vector order parameter aligned with
the field ($\vec{h}\parallel\mathbf{\hat{d}}$). In this case a Zeeman
splitting of the differential conductance peak, similar to that in
conventional superconductors (N-S junctions), should be observed. As the field
is rotated, the splitting would be suppressed and for a magnetic field
oriented parallel to the $\mathbf{b}^{\prime}$-axis there would be no Zeeman
effect (accompanied by the possibility of applying large fields that are not
paramagnetically limited). The disappearance of splitting even as the field is
being increased would constitute a clear signature of spin-triplet superconductivity.

As in the case of N-T junctions, we can envisage using the Zeeman response of
S-T junctions as a direct probe for spin-triplet order. If for instance, a
field is applied along the $\mathbf{b}^{\prime}$-axis of \textrm{(TMTSF)}%
$_{2}$\textrm{PF}$_{6}$, we predict a Zeeman splitting of the main
differential conductance peak. This would constitute a clear sign of
unconventional superconductivity since such an effect does not take place for
standard BCS superconductors (S-S junctions). The $\mathbf{b}^{\prime}$
direction is the one on which the upper critical field is not paramagnetically
limited, so relatively large fields could be applied in order to obtain a
clear signal.

Our considerations could be extended to the case of the layered compounds
believed to be triplet superconductors\cite{ishiguro2002plus}. Among them,
\textrm{Sr}$_{2}$\textrm{RuO}$_{4}$ is the best studied so far, but only few
tunneling experiments were
performed\cite{jin2000,laube2000,mao2001,sumiyama2002,upward2002}, and none so
far with good resolution in the presence of an applied external magnetic field
to observe Zeeman effects. One of the conspicuous features observed in some of
these experiments is the presence of a zero-bias anomaly in the differential
conductance. Its explanation is still a matter of debate, but seems to require
extended contacts and sign changing order parameters. To include the effect of
`zero energy states' at the interfaces would require extensions to our scheme,
possibly incorporating certain aspects of those calculations already done for
planar junctions\cite{yamashiro1997,honerkamp1998,sengupta2001,asano2003}. Our
general findings about the effect of magnetic fields should however apply,
since they refer to features to be measured at voltages of the order of the
superconducting gap.

Summarizing, the point-contact tunneling involving unconventional
superconductors with spin-triplet pairing displays interesting characteristic
features. In particular the Zeeman response to an external magnetic field is
such that it may allow for the identification of triplet phases and might be
relevant for future experiments. The prediction of a truncated sub-gap
structure in S-T junctions is also very interesting. For future extensions of
this work, different effects (finite temperature, contact potentials in the
junction, spin-flip tunneling, etc.) can be easily incorporated to our calculations.

\bigskip

We thank \O . Fischer, M. Eskildsen, M. Kugler and G. Levy for discussions
about the practical aspects of tunneling experiments, and Y. Maeno and M.
Sigrist for discussions about \textrm{Sr}$_{2}$\textrm{RuO}$_{4}$. We also
thank I.~\v{Z}uti\'{c} for pointing out Ref.~\cite{sengupta2001}. This work
was supported by the Swiss National Science Foundation through MaNEP.


\begin{thebibliography}{999999999999999999999999999999999999999999999999999999999999999999999999999999999999}             %
\expandafter\ifx\csname natexlab\endcsname\relax

\fi \expandafter\ifx\csname bibnamefont\endcsname\relax 

\fi \expandafter\ifx\csname bibfnamefont\endcsname\relax 

\fi \expandafter\ifx\csname citenamefont\endcsname\relax 

\fi \expandafter\ifx\csname url\endcsname\relax 

\fi \expandafter\ifx\csname urlprefix\endcsname\relax

\fi \providecommand{\bibinfo}[2]{#2} \providecommand{\eprint}[2][]{\url{#2}}

\bibitem[Sigrist and Ueda(1991)]{sigrist1991}%
\bibinfo{author}{\bibfnamefont{M.}~\bibnamefont{Sigrist}} and
\bibinfo{author}{\bibfnamefont{K.}~\bibnamefont{Ueda}},
\bibinfo{journal}{Rev. Mod. Phys.} \textbf{\bibinfo{volume}{63}},
\bibinfo{pages}{239} (\bibinfo{year}{1991}).

\bibitem[Rice and Sigrist(1995)]{rice1995}%
\bibinfo{author}{\bibfnamefont{T.~M.} \bibnamefont{Rice}} and
\bibinfo{author}{\bibfnamefont{M.}~\bibnamefont{Sigrist}},
\bibinfo{journal}{J. Phys.: Condens. Matter} \textbf{\bibinfo{volume}{7}},
\bibinfo{pages}{L643} (\bibinfo{year}{1995}).

\bibitem[Mackenzie and Maeno(2003)]{mackenzie2003}%
\bibinfo{author}{\bibfnamefont{A.~P.} \bibnamefont{Mackenzie}} and
\bibinfo{author}{\bibfnamefont{Y.}~\bibnamefont{Maeno}},
\bibinfo{journal}{Rev. Mod. Phys.} \textbf{\bibinfo{volume}{75}},
\bibinfo{pages}{657} (\bibinfo{year}{2003}).

\bibitem[Ishiguro(2002)]{ishiguro2002plus}%
\bibinfo{author}{\bibfnamefont{T.}~\bibnamefont{Ishiguro}}, in
\emph{\bibinfo{booktitle}{High Magnetic Fields}}, edited by
\bibinfo{editor}{\bibfnamefont{C.}~\bibnamefont{Berthier}},
\bibinfo{editor}{\bibfnamefont{L.~P.} \bibnamefont{L\'evy}}, and
\bibinfo{editor}{\bibfnamefont{G.}~\bibnamefont{Martinez}}
(\bibinfo{publisher}{Springer-Verlag}, \bibinfo{address}{Heidelberg},
\bibinfo{year}{2002}). \bibinfo{note}{See also the contribution of V.~Mineev to the same volume}.

\bibitem[Lebed(1999)]{lebed1999}%
\bibinfo{author}{\bibfnamefont{A.~G.} \bibnamefont{Lebed}},
\bibinfo{journal}{Phys. Rev. B} \textbf{\bibinfo{volume}{59}},
\bibinfo{pages}{R721} (\bibinfo{year}{1999}).

\bibitem[Blonder et~al.(1982)Blonder, Tinkham, and Klapwijk]{blonder1982}%
\bibinfo{author}{\bibfnamefont{G.~E.} \bibnamefont{Blonder}},
\bibinfo{author}{\bibfnamefont{M.}~\bibnamefont{Tinkham}}, and
\bibinfo{author}{\bibfnamefont{T.~M.} \bibnamefont{Klapwijk}},
\bibinfo{journal}{Phys. Rev. B} \textbf{\bibinfo{volume}{25}},
\bibinfo{pages}{4515} (\bibinfo{year}{1982}).

\bibitem[Octavio et~al.(1983)Octavio, Tinkham, Blonder, and Klapwijk]%
{octavio1983}\bibinfo{author}{\bibfnamefont{M.}~\bibnamefont{Octavio}}
\textit{et al.}, \bibinfo{journal}{Phys. Rev. B} \textbf{\bibinfo{volume}{27}}%
, \bibinfo{pages}{6739} (\bibinfo{year}{1983}).

\bibitem[Cohen et~al.(1962)Cohen, Falicov, and Phillips]{cohen1962}%
\bibinfo{author}{\bibfnamefont{M.~H.} \bibnamefont{Cohen}},
\bibinfo{author}{\bibfnamefont{L.~M.} \bibnamefont{Falicov}}, and
\bibinfo{author}{\bibfnamefont{J.~C.}
\bibnamefont{Phillips}}, \bibinfo{journal}{Phys. Rev. Lett.}
\textbf{\bibinfo{volume}{8}}, \bibinfo{pages}{316} (\bibinfo{year}{1962}).

\bibitem[Wilkins(1969)]{wilkins1969}%
\bibinfo{author}{\bibfnamefont{J.~W.} \bibnamefont{Wilkins}}, in
\emph{\bibinfo{title}{Tunneling Phenomena in Solids}}
(\bibinfo{publisher}{Plenum Press}, \bibinfo{address}{New York}, \bibinfo{year}{1969}).

\bibitem[Cuevas et~al.(1996)Cuevas, Mart\'{\i}n-Rodero, and Yeyati]%
{cuevas1996}\bibinfo{author}{\bibfnamefont{J.~C.} \bibnamefont{Cuevas}},
\bibinfo{author}{\bibfnamefont{A.}~\bibnamefont{Mart\'{\i}n-Rodero}}, and
\bibinfo{author}{\bibfnamefont{A.~L.}
\bibnamefont{Yeyati}}, \bibinfo{journal}{Phys. Rev. B}
\textbf{\bibinfo{volume}{54}}, \bibinfo{pages}{7366} (\bibinfo{year}{1996}).

\bibitem[Keldysh(1965)]{keldysh1965}%
\bibinfo{author}{\bibfnamefont{L.~V.} \bibnamefont{Keldysh}},
\bibinfo{journal}{Sov. Phys. JETP} \textbf{\bibinfo{volume}{24}},
\bibinfo{pages}{1018} (\bibinfo{year}{1965}).

\bibitem[Sengupta et~al.(2001)Sengupta, {\v{Z}}uti\'{c}, Kwon, Yakovenko, and
{Das Sarma}]{sengupta2001}%
\bibinfo{author}{\bibfnamefont{K.}~\bibnamefont{Sengupta}} \textit{et al.},
\bibinfo{journal}{Phys. Rev. B} \textbf{\bibinfo{volume}{63}},
\bibinfo{pages}{144531} (\bibinfo{year}{2001}).

\bibitem[Lee et~al.(1997)Lee, Naughton, Danner, and Chaikin]{lee1997}%
\bibinfo{author}{\bibfnamefont{I.~J.} \bibnamefont{Lee}} \textit{et al.},
\bibinfo{journal}{Phys. Rev. Lett.} \textbf{\bibinfo{volume}{78}},
\bibinfo{pages}{3555} (\bibinfo{year}{1997}).

\bibitem[Lee et~al.(2000)Lee, Chaikin, and Naughton]{lee2000}%
\bibinfo{author}{\bibfnamefont{I.~J.} \bibnamefont{Lee}},
\bibinfo{author}{\bibfnamefont{P.~M.} \bibnamefont{Chaikin}}, and
\bibinfo{author}{\bibfnamefont{M.~J.}
\bibnamefont{Naughton}}, \bibinfo{journal}{Phys. Rev. B}
\textbf{\bibinfo{volume}{62}}, \bibinfo{pages}{R14669} (\bibinfo{year}{2000}).

\bibitem[Lee et~al.(2002{a})Lee, Chaikin, and Naughton]{lee2002a}%
\bibinfo{author}{\bibfnamefont{I.~J.} \bibnamefont{Lee}},
\bibinfo{author}{\bibfnamefont{P.~M.} \bibnamefont{Chaikin}}, and
\bibinfo{author}{\bibfnamefont{M.~J.}
\bibnamefont{Naughton}}, \bibinfo{journal}{Phys. Rev. B}
\textbf{\bibinfo{volume}{65}}, \bibinfo{pages}{180502(R)} (\bibinfo{year}{2002}{\natexlab{a}}).

\bibitem[Lee et~al.(2002{b})Lee, Chaikin, and Naughton]{lee2002b}%
\bibinfo{author}{\bibfnamefont{I.~J.} \bibnamefont{Lee}},
\bibinfo{author}{\bibfnamefont{P.~M.} \bibnamefont{Chaikin}}, and
\bibinfo{author}{\bibfnamefont{M.~J.}
\bibnamefont{Naughton}}, \bibinfo{journal}{Phys. Rev. Lett.}
\textbf{\bibinfo{volume}{88}}, \bibinfo{pages}{207002} (\bibinfo{year}{2002}{\natexlab{b}}).

\bibitem[Oh and Naughton(2004)]{oh2004}%
\bibinfo{author}{\bibfnamefont{J.~I.} \bibnamefont{Oh}} and
\bibinfo{author}{\bibfnamefont{M.~J.} \bibnamefont{Naughton}},
\bibinfo{journal}{Phys. Rev. Lett.} (\bibinfo{year}{2004}), \bibinfo{note}{[to be published]}.

\bibitem[Lebed et~al.(2000)Lebed, Machida, and Ozaki]{lebed2000}%
\bibinfo{author}{\bibfnamefont{A.~G.} \bibnamefont{Lebed}},
\bibinfo{author}{\bibfnamefont{K.}~\bibnamefont{Machida}}, and
\bibinfo{author}{\bibfnamefont{M.}~\bibnamefont{Ozaki}},
\bibinfo{journal}{Phys. Rev. B} \textbf{\bibinfo{volume}{62}},
\bibinfo{pages}{R795} (\bibinfo{year}{2000}).

\bibitem[Lee et~al.(2002{c})Lee, Brown, Clark, Strouse, Naughton, Kang, and
Chaikin]{lee2002c}\bibinfo{author}{\bibfnamefont{I.~J.} \bibnamefont{Lee}}
\textit{et al.}, \bibinfo{journal}{Phys. Rev. Lett.}
\textbf{\bibinfo{volume}{88}}, \bibinfo{pages}{017004} (\bibinfo{year}{2002}{\natexlab{c}}).

\bibitem[Tanuma et~al.(2002)Tanuma, Kuroki, Tanaka, Arita, Kashiwaya, and
Aoki]{tanuma2002}\bibinfo{author}{\bibfnamefont{Y.}~\bibnamefont{Tanuma}}
\textit{et al.}, \bibinfo{journal}{Phys. Rev. B} \textbf{\bibinfo{volume}{66}}%
, \bibinfo{pages}{094507} (\bibinfo{year}{2002}).

\bibitem[Jin et~al.(2000)Jin, Liu, Mao, and Maeno]{jin2000}%
\bibinfo{author}{\bibfnamefont{R.}~\bibnamefont{Jin}} \textit{et al.},
\bibinfo{journal}{Europhys. Lett.} \textbf{\bibinfo{volume}{51}},
\bibinfo{pages}{341} (\bibinfo{year}{2000}).

\bibitem[Laube et~al.(2000)Laube, Goll, v.~L{\"{o}}hneysen, Fogelstr{\"{o}}m,
and Lichtenberg]{laube2000}%
\bibinfo{author}{\bibfnamefont{F.}~\bibnamefont{Laube}} \textit{et al.},
\bibinfo{journal}{Phys. Rev. Lett.} \textbf{\bibinfo{volume}{84}},
\bibinfo{pages}{1595} (\bibinfo{year}{2000}).

\bibitem[Mao et~al.(2001)Mao, Nelson, Jin, Liu, and Maeno]{mao2001}%
\bibinfo{author}{\bibfnamefont{Z.~Q.} \bibnamefont{Mao}} \textit{et al.},
\bibinfo{journal}{Phys. Rev. Lett.} \textbf{\bibinfo{volume}{87}},
\bibinfo{pages}{037003} (\bibinfo{year}{2001}).

\bibitem[Upward et~al.(2002)Upward, Kouwenhoven, Morpurgo, Kikugawa, Mao, and
Maeno]{upward2002}\bibinfo{author}{\bibfnamefont{M.~D.} \bibnamefont{Upward}}
\textit{et al.}, \bibinfo{journal}{Phys. Rev. B} \textbf{\bibinfo{volume}{65}}%
, \bibinfo{pages}{220512} (\bibinfo{year}{2002}).

\bibitem[Sumiyama et~al.(2002)Sumiyama, Endo, Oda, Yoshida, Mukai, Ono, and
\={O}nuki]{sumiyama2002}%
\bibinfo{author}{\bibfnamefont{A.}~\bibnamefont{Sumiyama}} \textit{et al.},
\bibinfo{journal}{Physica C} \textbf{\bibinfo{volume}{367}},
\bibinfo{pages}{129} (\bibinfo{year}{2002}).

\bibitem[Yamashiro et~al.(1997)Yamashiro, Tanaka, and Kashiwaya]%
{yamashiro1997}\bibinfo{author}{\bibfnamefont{M.}~\bibnamefont{Yamashiro}},
\bibinfo{author}{\bibfnamefont{Y.}~\bibnamefont{Tanaka}}, and
\bibinfo{author}{\bibfnamefont{S.}~\bibnamefont{Kashiwaya}},
\bibinfo{journal}{Phys. Rev. B} \textbf{\bibinfo{volume}{56}},
\bibinfo{pages}{7847} (\bibinfo{year}{1997}).

\bibitem[Honerkamp and Sigrist(1998)]{honerkamp1998}%
\bibinfo{author}{\bibfnamefont{C.}~\bibnamefont{Honerkamp}} and
\bibinfo{author}{\bibfnamefont{M.}~\bibnamefont{Sigrist}},
\bibinfo{journal}{J. Low Temp. Phys} \textbf{\bibinfo{volume}{111}},
\bibinfo{pages}{895} (\bibinfo{year}{1998}).

\bibitem[Asano et~al.(2003)Asano, Tanaka, Sigrist, and Kashiwaya]%
{asano2003}\bibinfo{author}{\bibfnamefont{Y.}~\bibnamefont{Asano}} \textit{et
al.}, \bibinfo{journal}{Phys. Rev. B} \textbf{\bibinfo{volume}{67}},
\bibinfo{pages}{184505} (\bibinfo{year}{2003}).
\end{thebibliography}

\end{document}